\documentstyle[12pt]{article}
\def\MB#1{\mbox{{\scriptsize{\rm#1}}}}
\newcommand{\arctanh}{\mathop{\rm arctanh}\nolimits}

\begin{document}

\title{A proposal for a new type of thin-film field-emission display
by edge breakdown of MIS structure}

\author{
V.N. Konopsky
}

\date{\it
Institute of Spectroscopy, Russian Academy of Sciences, \\
Troitsk, Moscow region, 142092, Russia.\\
(e-mail: konopsky@isan.troitsk.ru)
}

\maketitle

\begin{abstract}
A new type of field emission display~(FED)
based on an edge-enhance electron emission from
metal-insulator-semiconductor~(MIS) thin film structure is proposed.
The electrons produced by an  avalanche breakdown in the
semiconductor near the edge of a top metal electrode are initially
injected to the thin film of an insulator with a negative electron
affinity~(NEA), and then are injected into vacuum in proximity to the
top electrode edge. The condition for the deep-depletition breakdown
near the edge of the top metal electrode is analytically found in
terms of ratio of the insulator thickness to the maximum (breakdown)
width of the semiconductor depletition region: this ratio should be
less than $2/(3\pi -2)\simeq 0.27$.   The influence of a neighboring
metal electrode and an electrode thickness on this condition  are
analyzed.  Different practical schemes of the proposed display with a
special reference to $\rm M/CaF_2/Si$ structure are considered.
\end{abstract}

\section{Introduction}
Flat panel field emission displays have the potential to be a low
cost, high performance alternative to the currently dominant cathode
ray tube and liquid crystal displays. The first major problem in FEDs
is the development of a reliable and efficient cold cathode electron
emitter.  Current FED prototypes use sharp metal or semiconductor
tips
as field emitters~\cite{Spindt}, that requires expensive
lithography and other difficult fabrication processes. Besides, the
control voltage for such  tip emitters is rather high (about 100~V).
Some researchers use diamond-like films that contain nanoscale
crystalline structure as the electron source~\cite{Zhu_art}, or use
diamond and other coatings of the sharp tips to improve the emission
properties of the tips~\cite{Liu, Konopsky}.

In the present paper we propose another type of field emitter, where
electrons are produced by an avalanche breakdown in the semiconductor
near the edge of the top metal electrode in the MIS structure.

The plan of this paper is as follows: in Section~2 we describe
the proposed display, based on the edge breakdown of the MIS
structure. In Section~3 we discuss the condition under which edge
breakdown in the MIS structure takes place.  In Section~4 we
estimate the influence of a neighboring electrode on the edge
breakdown condition and in Section~5 we discuss and summarize our
results. In Appendix we briefly estimate the influence of an
electrode thickness on  the edge breakdown condition.

\section{Field emission display based on edge breakdown of MIS
structure}
The display proposed is schematically shown in Figure~1(a).
Glass substrate~1 with conductive metal column lines~2 is
coated with a thin semiconductor layer~3, which contains low
p-doped column lines~4 that coincide with the metal ones. The
film of the insulator with NEA~5
is grown on semiconductor, and conductive metal row lines~6
are deposited on the insulator film. Above the MIS structure the
fluorescent screen~7 is located. The anode (fluorescent screen) and
cathode (MIS structure) regions are separated by a vacuum space~8.

When a positive voltage pulse of duration which is short compared to
the time constant of thermal generation of minority carriers
(electrons in our case) is applied to the top electrode of the
MIS structure, no inversion layer can form. Thus, a large potential
drop across the semiconductor will take place. If the amplitude of
the pulse voltage is increased,
band bending reaches large values where minority carriers are
generated by nonthermal effects (by avalanche in our case). The
amplitude of the pulse voltage, at which avalanche occurs we
designate by $V_{\MB{break}}$.

When we apply a pulse voltage $V_{\MB{control}}$, which amplitude is
less then $V_{\MB{break}}$, but more than $V_{\MB{break}}/2$
to one row line, and
simultaneously apply such a pulse voltage with another polarity to one
column line, the avalanche breakdown in the semiconductor will take
place at the intersection of these lines. Under condition, which will
be pointed out below, the breakdown will take place near the edges
of metal lines. Inasmuch as velocities of the avalanche electrons
at the semiconductor-insulator interface are directed in an arbitrary
way, it is clear that at least a portion of the avalanche electrons,
ballistically passing through the thin film of the insulator, will be
extracted with a good efficiency (due to NEA of the insulator) into
vacuum and will not impinge on the top metal layer. Then
this portion of electrons will be accelerated by screen voltage
$V_{\MB{screen}}$ ($\sim +100\div +1000$~V) and will
be hit to the fluorescent screen as it is shown in Figure~1(b).

\section{Edge breakdown condition}
The deep-depletition breakdown of MOS capacitors was first reported
by Goetsberger and Nicollian~\cite{Goetzberger1,Goetzberger2}, who
experimentally investigated doping conditions under which uniform
avalanche takes place. Later a ``universal and normalized'' criterion
for ``field uniformity'' in MOS capacitors was offered in the form
$d/W_{\MB{max}}>0.3$ (where $d$ is the insulator thickness and
$W_{\MB{max}}$ is the maximum (breakdown) width of the semiconductor
depletition region)~\cite{Rusu,Sze_book}. This criterion was
suggested in~\cite{Rusu} based on computer-calculated values of the
field distribution in MOS capacitors.

In this section we obtain this criterion in an analytical form, with
an emphasis on physical explanation of the result, that helps us to
analyze operating conditions and an ultimate resolution of the
display proposed.

For this purpose let us at first consider a simple two-dimensional
model of the electric field near the edge of the metal plate (see
Figure~2(a)). Here $A_1-A_2$ and $A_2-A_3-A_4$ are conductive
plates. Let one plate have the potential $V$, and the other plate
have the zero potential.

We will use a method of conformal transformations
(see~\cite{Panovsky} or any textbook in this field) to find an
electrical field distribution in such a system. The
Schwarz-Chrisoffel transform
\begin{equation}
z={\i h\over \pi}\left[ 2\sqrt{\omega-1}-\i\ln\left(
{1-\i\sqrt{\omega-1}\over 1+\i\sqrt{\omega-1}}\right) \right]
\label{1}
\end{equation}
relates the upper half $\omega$ plane (Figure~2(b)) to
the interior of
the region $A_1-A_2-A_3-A_4$ of the $z$ plane
(Figure~2(a))~\cite{Korn}.
Thus, half of the real axis $\Re{(\omega)} >0$ is at potential $V$
while $\Re{(\omega)} <0$ is at zero potential. The potential of the
electric field in the $\omega$ plane is the real part of the complex
potential given by the analytical function
\begin{equation}
F=\varphi +\i\psi ={V\over \i\pi}\ln\omega \; .
\label{2}
\end{equation}
The electric field in the $z$ plane is
\begin{equation}
{d F\over d z} = {\partial\varphi\over \partial
x} - \i {\partial\varphi\over \partial y} = E_x-\i E_y = {d F\over
d \omega}{1\over d z/ d \omega} \; .
\label{3}
\end{equation}
Performing the differentiation with respect to $\omega$ in~(\ref{1})
and~(\ref{2})
\begin{equation}
{d z\over d\omega}={\i h\over
\pi}{\sqrt{\omega-1}\over\omega}
\label{4}
\end{equation}
\begin{equation}
{d F\over d\omega}={V\over \i\pi\omega}
\label{5}
\end{equation}
we get
\begin{equation}
E_x-\i E_y = -{V\over h\sqrt{\omega-1}} \; .
\label{6}
\end{equation}

Our main interest is the field near $z=0$ ($\omega=1$).
Expanding logarithm in~(\ref{1}) in a Taylor series
at $\omega=1$ and holding
the first two terms, we obtain
\begin{equation}
\sqrt{\omega-1} = \left( {3\pi\over 2h\i}z \right)^{1/3}
\label{7}
\end{equation}
and from (\ref{6}) we have
\begin{equation}
E_x-\i E_y = -{V\over h\left(3\pi z/2h\i
\right)^{1/3}}
\label{71}
\end{equation}
or
\begin{equation}
\left| E\right|={V\over h}{\left( 2h/3\pi\right)^{1/3}
\over\left( x^2+y^2\right)^{1/6}}  \; .
\label{8}
\end{equation}
We may define
\begin{equation}
z_{\MB{cr}}={2h\over 3\pi} \simeq {h\over 4.71}
\label{9}
\end{equation}
as a ``critical'' distance from the edge at which field strength is
still higher than $V/h$ --- the field strength between the plates far
from the edge.

Now the condition for edge breakdown is rather obvious: if
insulator thickness is less than $z_{\MB{cr}}$ (as is shown in
Figure~3) the breakdown takes place near the edge. But if
$d>z_{\MB{cr}}$ the uniform breakdown occurs. The value
$d+W_{\MB{max}}$ in the real MIS structure plays the role of the
distance $h$ between the conductive plates in our model
consideration. (The exception is the case when the distance $L$
(see Figure~1(b)) between the real metal plates is less than
$d+W_{\MB{max}}$. In this case $h=L$.)  From the equations
\begin{equation}
h=d+W_{\MB{max}}
\label{10}
\end{equation}
\begin{equation}
d<z_{\MB{cr}}={2h\over 3\pi}
\label{11}
\end{equation}
we may obtain the criteria for
the deep-depletition edge breakdown in
terms of ratio of the insulator thickness to the maximum (breakdown)
width of the semiconductor depletition region:
\begin{equation}
{d\over W_{\MB{max}}}<{2\over 3\pi -2}\simeq 0.27\; .
\label{12}
\end{equation}

One can see that the value obtained ($0.27$) is consistent with the
value $0.3$ obtained in the work~\cite{Rusu}
by numerical computer-aided calculations
for particular system $\rm M/SiO_2/Si$.

The model
considered  is applicable to real systems subject to the condition
that the radius of the edge curvature of the top electrode is less
than insulator thickness.

\section{Breakdown condition for double-edge structure}
In this Section we apply analogous approach to find breakdown
condition for the system illustrated in Figure~4(a). It is
necessary for several reasons: first, one may like to
initiate electron emission into a slit in the top electrode. Second,
it may be convenient to apply a voltage simultaneously to a set of
neighboring row and column lines for more stable emission. And third,
minimal distance between top electrodes defines the ultimate
resolution of the proposed display.

The transformation
\begin{equation}
z={2\i\over \pi}\left[
s\arctanh{\left(\omega\over\sqrt{\omega^2-\lambda^2}\right)}
+h\arctan{\left({h\over s}{\omega\over\sqrt{\omega^2-\lambda^2}}
\right)}
\right]
\label{13}
\end{equation}
where
\begin{equation}
\lambda=\sqrt{1+{h^2\over s^2}}
\label{14}
\end{equation}
relates the upper half $\omega$ plane (Figure~4(b)) to
the interior of
the region $A_1-A_2-A_3-A_4-A_5-A_6$ of the $z$ plane
(Figure~4(a))~\cite{Lavr}.
Thus, parts of the real axis $\Re{(\omega)} <-1$ and $\Re{(\omega)}
>1$ are at potential $V$ while part $-1<\Re{(\omega)} <1$ is at zero
potential. The potential of the electric field in the $\omega$ plane
is the real part of the complex potential given by the analytical
function
\begin{equation}
F={V\over \i\pi}\ln{(w-1)} - {V\over \i\pi}\ln{(w+1)}\; .
\label{15}
\end{equation}
Using the differentials with respect to $\omega$ of (\ref{13})
\begin{equation}
{d z\over d\omega}={2\i s\over
\pi}{\sqrt{\omega^2-\lambda^2}\over(\omega^2-1)}
\label{16}
\end{equation}
and (\ref{15})
\begin{equation}
{d F\over d\omega}={2V\over \i\pi(\omega^2-1)}
\label{17}
\end{equation}
we may find the electric field in the $z$ plane from (\ref{3}).
\begin{equation}
E_x-\i E_y = -{V\over s\sqrt{\omega^2-\lambda^2}} \; .
\label{18}
\end{equation}
Our concern is the field near $z=\i h\pm s$  ($\omega=\pm\lambda$)
points.
Expanding arctanh and arctan in~(\ref{13}) in a power series at
$\omega =\lambda$ and holding the first two terms, we have
\begin{equation}
\sqrt{\omega^2-\lambda^2} = {h\over s}\left[ {3\pi \over 2h\i}(z+h-\i
s)\left(1+{h^2\over s^2}\right)^2 \right]^{1/3} \; .
\label{19}
\end{equation}
and from (\ref{18}) we have
\begin{equation}
E_x-\i E_y = -{V\over h\left[ (3\pi/2h\i)(z+h-\i
s)\left(1+{h^2/s^2}\right)^2 \right]^{1/3}} \; .
\label{20}
\end{equation}
Now the ``critical'' distance from the edge
at which field strength is still higher than
$V/h$ is
\begin{equation}
z_{\MB{cr2}}={2h\over 3\pi \left( 1+h^2/s^2\right)^2 }\; .
\label{21}
\end{equation}
One can see that this value is coincident with~(\ref{9}) when
$s\gg h$.

For example, if $s=h$
so $z_{\MB{cr2}}$ is
\begin{equation}
z_{\MB{cr2}}={z_{\MB{cr}}\over 4}={h\over 6\pi}\; .
\label{22}
\end{equation}
One can see that
the distance $2s$ between the top electrodes should be more than twice
larger than $h=d+W_{\MB{max}}$ to avoid drastic decrease of
$z_{\MB{cr2}}$.

\section{Discussion and summary}
Firstly, let us mention the kind of insulator that may be used in
this display. At least three kinds of insulators are appropriate for
our purpose: hydrogen-terminated diamond~(111), LiF and $\rm CaF_2$.
Diamond has long attracted considerable attention as a cold
cathode for FED due to its NEA and robust mechanical and chemical
properties.

LiF has the largest  NEA of any solid. The NEA of LiF
crystal is $-2.7$~eV~\cite{Lapiano}. The possibility of epitaxial
grown of LiF films on the Ge was also reported in~\cite{Lapiano}.

But we assume that the $\rm CaF_2/Si$ system is the best choice.
The reasons for this are as follows:
\begin{enumerate}
\item $\rm CaF_2$ has a small but negative electron
affinity~\cite{Poole}.
\item A very attractive property of $\rm CaF_2$ is similarity of
its crystal structure to Si: $\rm CaF_2$ has cubic m3m structure with
only 0.6~\% of lattice mismatch with Si(111) at room temperature.
This fact makes it possible to grow perfect single crystal
films of $\rm CaF_2$ on silicon by molecular beam epitaxy
technique~\cite{Batstone, Rieger}.  The crystalline quality, chemical
stability and electrical characteristics of $\rm CaF_2$ films grown
on Si(111)  may be further improved by rapid thermal
annealing~\cite{Phillips}.
\item Barrier height between the conductive band minimum of Si and
the conductive band minimum of $\rm CaF_2$ is only $\sim
2.2$~eV~\cite{Rieger}.
\item An additional attractive property
of $\rm CaF_2$ is anomalously large low-energy electron escape depth
of the order of 260~nm~\cite{Quiniou}.
\item And lastly, the good
emission properties of the $\rm CaF_2/Si$ structure has recently
been demonstrated in the experiment~\cite{Konopsky}.
\end{enumerate}
The thickness of the insulator film about $d\sim 10$~nm seems to be
nearly optimum because it is large enough to avoid direct tunneling
of electrons from semiconductor to metal, but it is small enough for
ballistic passing of electrons through insulator film.

So, if one has 10~nm thick $\rm CaF_2$ film on the $1~\mu$m thick Si
layer with p-dopant concentration $n\sim 10^{16}$cm${^{-3}}$
($W_{\MB{max}}\sim 1~\mu$m~\cite{Sze_book}), one may be sure that
when  pulse voltages with the amplitude
$V_{\MB{control}}\sim\pm 10$~V are
applied to one row and one column lines an avalanche breakdown will
occur at the edges of metal lines ($d/W_{\MB{max}}\sim 0.01$).
The control voltage may be still decreased if
the distance between metal electrodes $L$
will be less than
$d+W_{\MB{max}}$
(but $L$ must always be more than $3\pi d/2$). The avalanche
electrons will be injected into insulator at the $\alpha - \beta$
line, which shown in Figure~3. From geometrical considerations,
up to several tens percent of electrons may be injected into vacuum.

To summarize, we have proposed a simple design of the field-emission
display based on the deep-depletition edge avalanche breakdown of the
MIS structure. The control voltage of this display may be as small as
tens volts and its ultimate resolution may reach several microns.

\section*{Appendix}
\appendix
Here we mention the influence of the top electrode thickness on the
edge breakdown condition.
To avoid complicated expressions, we give only a simple estimation of
the effect by solving the problem presented in Figure~5(a), where
top electrode has zero thickness.  The equation
\begin{equation}
z={h\over \pi}(\omega-\ln{(w)}-1)
\label{A1}
\end{equation}
relates the upper half $\omega$ plane (Figure~5(b)) to
the interior
of the region $A_1-A_2-A_3-A_4$ of the $z$ plane
(Figure~5(a)).
The potential of the
electric field in the $\omega$ plane is the real part of the complex
potential given by the analytical function
\begin{equation}
F={V\over \i\pi}\ln\omega \; .
\label{A2}
\end{equation}
The electric field in the $z$ plane is
\begin{equation}
{d F\over d z} =
{d F\over
d\omega}{1\over d z/d\omega} \; .
\label{A3}
\end{equation}
Performing the differentiation with respect to $\omega$ in~(\ref{A1})
and~(\ref{A2})
\begin{equation}
{d z\over d\omega}={h\over
\pi}{{\omega-1}\over\omega}
\label{A4}
\end{equation}
\begin{equation}
{d F\over d\omega}={V\over \i\pi\omega}
\label{A5}
\end{equation}
we get
\begin{equation}
E_x-\i E_y = {V\over \i h(\omega-1)} \; .
\label{A6}
\end{equation}

Our main interest is the field near $z=0$ ($\omega=1$).
Expanding logarithm in~(\ref{A1}) in a Taylor series
at $\omega=1$ and holding
the first two terms, we obtain
\begin{equation}
{\omega-1} = \left( -{2\pi\over h}z \right)^{1/2}
\label{A7}
\end{equation}
and from (\ref{6}) we have
\begin{equation}
E_x-\i E_y = -{V\over h\left(2\pi z/h
\right)^{1/2}}\; .
\label{A8}
\end{equation}
One can see that the ``critical'' distance in this case is:
\begin{equation}
z_{\MB{cr0}}={h\over 2\pi }\; .
\label{A9}
\end{equation}
And the edge breakdown condition is
\begin{equation}
{d\over W_{\MB{max}}}<{1\over 2\pi -1}\simeq 0.19\; .
\label{A10}
\end{equation}
This edge breakdown condition should be used when top electrode
thickness
is less than any other dimensions in the system. Essentially it
means that top electrode thickness should be less than thickness of
the insulator film.

\newpage{}

\section*{
Figures' Captions
}
\ \\
{\bf Figure~1:}
{ Sectional view~(a) and side view~(b) of the
FED based on edge breakdown of MIS
structure.
1 --- glass substrate,
2 --- metal column lines,
3 --- semiconductor layer,
4 --- p-doped column lines,
5 --- insulator with NEA,
6 --- metal row lines,
7 --- fluorescent screen,
8 --- vacuum spacing.
}                                             \\[0.5cm]
{\bf Figure~2:}
{ Plane $z$~(a) and plane $\omega$~(b)
related by the equation~(\ref{1})
}                                             \\[0.5cm]
{\bf Figure~3:}
{ Distance
$z_{\MB{cr}}$ and
electron emission area
$\alpha - \beta$ near the edge.
}                                             \\[0.5cm]
{\bf Figure~4:}
{ Plane $z$~(a) and plane $\omega$~(b)
related by the equation~(\ref{13})
}                                             \\[0.5cm]
{\bf Figure~5:}
{ Plane $z$~(a) and plane $\omega$~(b)
related by the equation~(\ref{A1}) (Appendix)
}

\end{document}